\documentclass[manuscript]{acmart}
\AtBeginDocument{%
  \providecommand\BibTeX{{%
    \normalfont B\kern-0.5em{\scshape i\kern-0.25em b}\kern-0.8em\TeX}}}

\setcopyright{acmcopyright}
\copyrightyear{2018}
\acmYear{2018}
\acmDOI{XXXXXXX.XXXXXXX}

\acmConference[The ACM CHI conference]{Make sure to enter the correct conference title from your rights confirmation emai}{May 11-16, 2024}{Honolulu, USA}
\acmBooktitle{CHI} 
\acmPrice{15.00}
\acmISBN{978-1-4503-XXXX-X/18/06}

\acmSubmissionID{9802}

\begin{document}

\title{Prompt Your Mind: Refine Personalized Text Prompts within Your Mind}

\author{Guinan Su}
\email{guinansu33@gmail.com}
\affiliation{%
  \institution{Tencent Data Platform}
  \country{China}
}

\author{Yanwu Yang}
\affiliation{%
  \institution{Harbin Institute of Technology at Shenzhen}
  \country{China}
% \email{yangyanwu1111@gmail.com}
}
\author{Jie Guo}
\affiliation{%
  \institution{Peng Cheng Laboratory}
  \country{China}
}

% \affiliation{%

% \affiliation{%

% \affiliation{%

% \affiliation{%

% \affiliation{%

%\renewcommand{\shortauthors}{Anonymous, et al.}

\begin{abstract}
Large language models (LLMs) have demonstrated remarkable potential in natural language understanding and generation, making them valuable tools for enhancing conversational interactions. However, LLMs encounter challenges such as lacking multi-step reasoning capabilities, and heavy reliance on prompts. In this regard, we introduce a prompt-refinement system named PromptMind, also known as "Prompt Your Mind", to provide an automated solution for generating contextually relevant prompts during conversations. PromptMind enhances the overall interaction between humans and chatbots through an automatic prompt suggestion and an automatic prompt refinement. To assess the effectiveness of PromptMind, we designed three interaction tasks to evaluate emotional support, advice acquisition, and task-oriented interactions during human-chatbot interactions. The results demonstrated that PromptMind reduced mental demands during interactions and fostered enhanced performance and social connections between users and chatbots. In summary, our findings indicate that PromptMind acts as a bridge, facilitating smoother information exchange and enhancing the usability of chatbot interactions.

\end{abstract}

\begin{CCSXML}
<ccs2012>
   <concept>
       <concept_id>10003120.10003121.10003124.10010870</concept_id>
       <concept_desc>Human-centered computing~Natural language interfaces</concept_desc>
       <concept_significance>500</concept_significance>
       </concept>
   <concept>
       <concept_id>10003120.10003121.10003124.10010868</concept_id>
       <concept_desc>Human-centered computing~Web-based interaction</concept_desc>
       <concept_significance>300</concept_significance>
       </concept>
   <concept>
       <concept_id>10003120.10003121.10003129.10010885</concept_id>
       <concept_desc>Human-centered computing~User interface management systems</concept_desc>
       <concept_significance>300</concept_significance>
       </concept>
 </ccs2012>
\end{CCSXML}

\ccsdesc[500]{Human-centered computing~Natural language interfaces}
\ccsdesc[300]{Human-centered computing~Web-based interaction}
\ccsdesc[300]{Human-centered computing~User interface management systems}


\keywords{Prompt, large language model, human-chatbot interaction, multi-turn conversation}

\begin{teaserfigure}
  \includegraphics[width=\textwidth]{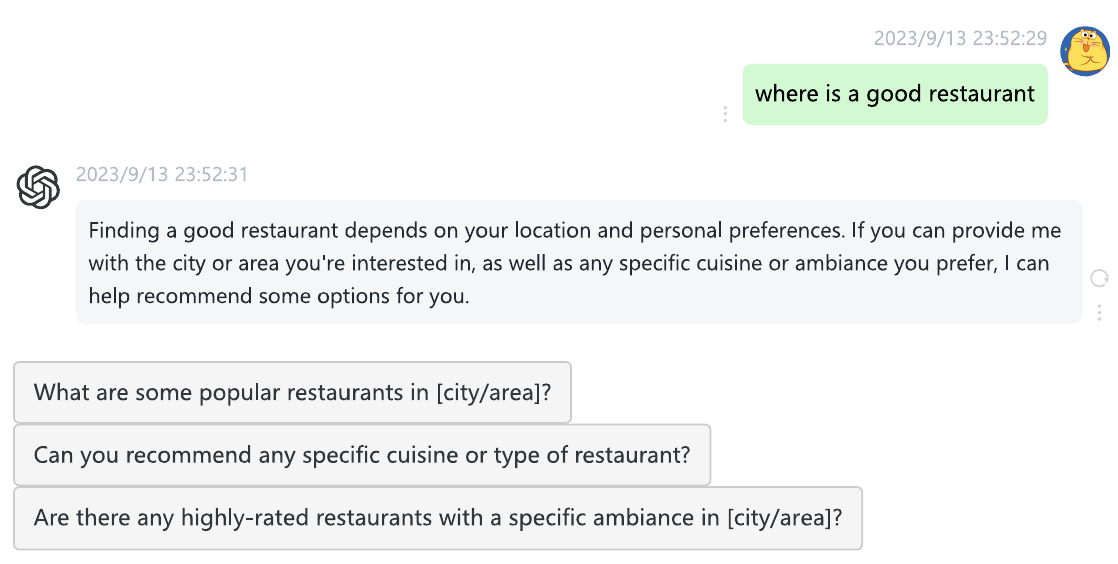}
  \caption{The PromptMind system UI. The ChatGPT-3.5 is used as the backend. PromptMind utilizes automatic prompt refinement to consider user feedback to improve prompt quality. Users can select from suggested prompts that are categorized as good cases for future improvement of the suggestion system. }
\end{teaserfigure}


\maketitle

\section{Introduction}
Large language models (LLMs), such as GPT and LLaMA, have revolutionized AI prototyping \cite{wu2022promptchainer,yang2020re,van2018prototyping,fitria2023artificial,kocaballi2023conversational}. Chatbots, in particular, have witnessed increasing applications across diverse domains like customer service, personal assistants, and online education \cite{xu2017new,liu2020cold,folstad2018sig,candello2018having,langevin2021heuristic,folstad2018sig,xiao2020tell}. These chatbot systems engage users in conversational dialogues, providing valuable information and support in various scenarios, including commercial transactions and emotional support\cite{chen2019jddc,zhao2021jddc,cheng2022improving,xie2019multi}.

The integration of natural language prompts with LLMs has emerged as a critical technique for tailoring these models to complex tasks. This synergy has shed light on natural language interactions, enabling advanced pre-trained LLMs to engage in fluent multi-turn conversations. This minimizes the hurdles related to data and programming skills, making conversational user experiences more accessible to non-AI experts \cite{zamfirescu2023conversation}. Users can enhance LLM outputs by adding prompts, which are textual instructions and examples of desired interactions. These prompts directly influence the model's output, expanding the potential of conversational user experiences.

However, despite the widespread excitement, several challenges persist in bridging the gap between chatbots and human understanding. LLMs sometimes struggle to accurately infer conversation topics and lack multi-step reasoning abilities \cite{li2022exploring,kojima2022large}. Misleading or inaccurate responses in multi-turn conversations can be detrimental. LLMs heavily rely on prompts, which can lead to inconsistent responses, causing user confusion. Moreover, even advanced chatbots may fall short of providing meaningful social benefits due to their lack of genuine consciousness. This absence of agency can hinder users from forming deep connections and limit their ability to offer empathetic and emotionally appropriate responses in certain situations \cite{folk2023can}.

%, due to the inconsistency inferring and potentially toxic responses. 
% In this regard, it is nontrivial to bridge the gap between chatbot understanding and human expressions. 
% Firstly, it is difficult for LLMs to infer what people are talking about and maintain consistency in multi-turn conversations without knowing background topic information \cite{li2022exploring}. 
% Moreover, Toxic responses, defined as offensive, insulting, or threatening statements that may pertain to issues of gender, politics, or race [22, 38], can be particularly harmful in the context of multiturn conversations. \cite{chen2023understanding}. 

% In this context, multi-turn interactions enable more complex tasks such as emotional counseling and comprehensive planning with up-down context learning. 

% In the past two years, social media platforms have witnessed an explosion of posts showing the results of lay peoples’ experimentation with LLMs for question answering, creative dialogue writing, writing code, and more. This excitement around LLMs and prompting is propelling a rapidly growing set of LLM-powered applications [23] and prompt design tools [3, 20, 32].
% develop enhancements to the model's context understanding, multi-step reasoning capabilities, and consistency

One approach to tackle the previously mentioned issue is to develop enhancements to the model's context understanding, multi-step reasoning capabilities, and consistency \cite{li2022exploring,zhang2018modeling,xu2021topic}. However, such studies rely on training with large-scale data, which are constrained by expressive computational costs and limitations related to scalability and generalizability. Another strategy involves prompt guidance. Prompts serve as potent tools for fostering interactive connections between humans and chatbots. Previous studies on prompt engineering have addressed these challenges by providing effective strategies for prompting \cite{liu2021makes,wang2023reprompt,brade2023promptify,white2023prompt}. Nevertheless, these strategies have limitations when it comes to offering personalized recommendations to users seeking specific topics and information. Moreover, they have yet to effectively establish social connections between chatbots and humans. Consequently, novice users lacking prior experience in prompt writing and familiarity with relevant keywords may still encounter difficulties in achieving their desired outcomes. 

To meet these issues, we explore three research questions in this paper:

\textit{
RQ1: How do laypersons perceive and engage with large language-based chatbots? }

\textit{RQ2: What techniques can be employed to automatically refine text prompts for improved chatbot responses?}

\textit{RQ3: How effective is the prompts refine approach in enhancing the quality of human-chatbot interactions?
}

For RQ1, we conducted formative user interviews with 14 users to gain a deeper understanding of the practices and challenges associated with prompting LLMs. The details about this interview study are shown in Section \ref{IS}. These interviews highlighted common issues, including misinterpretation, a lack of emotions and empathy, and deficiencies in multi-step reasoning, consistency, and personalization.

For the common issues mentioned in the interview study, we introduce Prompt Your Mind (PromptMind), an interactive system designed to streamline the iterative process of prompt exploration and refinement for LLMs (For RQ2). PromptMind facilitates the iterative workflow by incorporating various steps, including suggestions, and refined prompts. The automatic prompt suggestion step enables to provision of more comprehensive and contextually relevant responses. The automatic prompt refinement step utilizes user feedback to improve prompt quality. It provides three questions that align with the users' interests and potential queries as enhanced prompts for each round of dialogue to enhance the overall interaction between humans and chatbots. When users interact with chatbots, they can choose to continue by selecting one of the three questions offered by the system or by inputting their own questions based on their ideas.

To evaluate the effectiveness of PromptMind, we conducted a user study with 24 participants to compare PromptMind against the baseline tool, ChatGPT-3.5 (For RQ3). Participants consistently found PromptMind significantly more useful in assisting with the text-to-text generation workflow compared to the baseline tool. It also excelled in establishing social connections within the generated content. Additionally, users reported lower mental demands and frustration as well as enhanced efficiency and performance when using PromptMind. Importantly, these improvements were achieved without altering the underlying model. In summary, our work contributes the following:

\begin{itemize}
    \item An analysis of user perceptions of large language-based chatbots through an interview study ($n = 14$).
    \item The design and implementation of PromptMind, An LLM-based prompt suggestion engine that aids users in multi-turn conversation question prompting by providing personalized question recommendations.
    \item The results of a user study involving 24 participants demonstrate the efficacy of the PromptMind system and its superiority in providing meaningful social benefits and improving information seeking for users.
\end{itemize}

% and PromptMind suggests different ideas to extend the subject. They can further steer the subject suggestions to explore alternative ideas. Next, users input a short style description; Promptify then extends relevant modifiers 2 to articulate the style. The prompt suggestion engine can generate sophisticated prompts from users’ brief input, leading to visually appealing images. Promptify visualizes the generated images on an image layout and clustering interface, allowing users to flexibly organize and browse the generated collections.
%institutional email address
%%%related work
\section{Background and related works}
% Early language models, such as Long Short-Term Memory (LSTM) \cite{hochreiter1997long} and gated Recurrent Unit (GRU) \cite{chung2014empirical}, were constructed using different types of deep neural networks. However, with the emergence of the Transformer architecture, language models like BERT \cite{devlin2018bert} and GPT \cite{radford2018improving} have gained increasing popularity in Natural Language Generation (NLG) tasks, such as machine translation \cite{weng2020acquiring} and the generation of lengthy texts \cite{qu2020text}. These models are trained on extensive datasets to acquire a comprehensive understanding of text and the ability to generate long texts. The attention mechanism within the Transformer structure plays a pivotal role in the success of these pre-trained models. Consequently, large language models have become state-of-the-art AI technologies for various applications. These models, pre-trained on vast corpora, contain a wealth of information, allowing for easy customization during runtime without the need for re-training, enabling them to handle new tasks beyond text continuation. Based on this, the chatbots leveraging LLMs are proposed to handle conversation tasks.

Among the LLMs, ChatGPT-3.5 is a cutting-edge natural language processing model developed by OpenAI, designed to facilitate human-like conversations and interactions. GPT-3.5 is built upon advanced machine learning algorithms and extensive training data, which enable it to provide users with a seamless and engaging conversational experience.
It has been fine-tuned to handle a wide range of topics and adapt to various conversational styles, making it an ideal tool for applications such as customer support, virtual assistants, and chatbot development \cite{hill2023chat,surameery2023use,biswas2023role}.
Through continuous learning and improvement, GPT-3.5 strives to provide accurate, relevant, and empathetic responses to user queries, ensuring a satisfying and valuable user experience. Its impressive capabilities have garnered significant attention and recognition in the AI community, paving the way for future advancements in the field of natural language processing.
In addition, as part of Meta's commitment to open science, Meta has publicly released LLaMA \cite{touvron2023llama} (Large Language Model Meta AI), a state-of-the-art foundational language model designed to help researchers advance their work in this subfield of AI. Smaller, more performant models like LLaMA enable broader access for the research community, especially for those who may not have access to extensive infrastructure, further democratizing the field. Training smaller foundation models like LLaMA is advantageous in the large language model space, as they require significantly less computing power and resources. This makes it easier to test new approaches, validate others' work, and adapt them to specific domains. Llama, with its 7 billion parameters, offers an efficient and scalable solution, making it ideal for fine-tuning the automatic prompt suggestion model. In this study, we use the 7B-LLaMA model for fine-tuning prompt models.

Based on these LLMs, Chatbots are proposed to handle conversation tasks. In particular, chatbots engaging in multi-turn conversations have emerged as powerful tools for HCI researchers to investigate lingual interactions. To activate the desired functionality, users provide natural language instructions or prompts \cite{lu2021fantastically,liu2023pre}, tailored to the task at hand. Prompts are typically composed of a task description and/or several canonical examples. \cite{brown2020language} showed that Prompts are surprisingly effective at modulating a frozen LLM’s behavior through text prompts. Rather than requiring a separate copy of the model for each downstream task, a single generalist model can simultaneously serve many different tasks. \cite{lester2021power} further explores prompt tuning to perform specific downstream tasks. \cite{jia2022visual} introduces a visual prompt tuning for vision models. \cite{zhou2022learning} models a prompt's context words with learnable vectors for vision-language models. However, there still remains a lack of prompt tuning studies on multi-turn conversations. Multiple turns in a conversation can be used to understand the user information need more effectively and serve as important tools for information seeking. For instance, while searching for a new smartphone, the user and the machine could discuss various features and options in multiple turns. However, multi-turn natural language understanding still remains extremely challenging, requiring the system to comprehend the conversation context and reply in an informative and coincident manner \cite{cui2020mutual,aliannejadi2020harnessing,xu2021topic,zhang2018modeling}. Moreover, few studies have investigated user interactions with a machine to help the system understand their information needs.

\section{Interview Study}
\label{IS}

To understand current practices and limitations of how users of ChatGPT interact with the chatbot. We first conducted an online interview study. For our user studies, we manually built a website based on OPENAI API \cite{openai_gpt35turbo} for our participants. Therefore, we can save the generated conversation through the website backend. We took several API keys due to the request limit of ChatGPT3.5, which is necessary for real-time user interaction.

\subsection{Topic of Conversation }
% We have randomly selected 75 specific sample topics, divided into three categories: Emotional Support, Advice Acquisition, and Task-Oriented. We display the keywords of the topics in Figure \ref{words}.
% Emotional support topics focus on providing comfort and guidance during challenging times, such as coping with grief, managing stress, or maintaining a positive mindset during unemployment. Advice Acquisition topics revolve around seeking suggestions and recommendations for various situations, like planning a trip, improving time management skills, or adopting a healthier diet. Task-oriented topics involve working together with the chatbot to complete specific tasks or projects, such as planning a weekly fitness routine, organizing a charity event, or creating a budget for a home renovation.
% Participants can choose from the given sample topics or select their own within these three categories. We encourage participants to pick a topic that interests them and aligns with their goals or passions. This approach will help ensure a more engaging and productive discussion with the chatbot.

\begin{figure}
\includegraphics[width=0.8\linewidth]{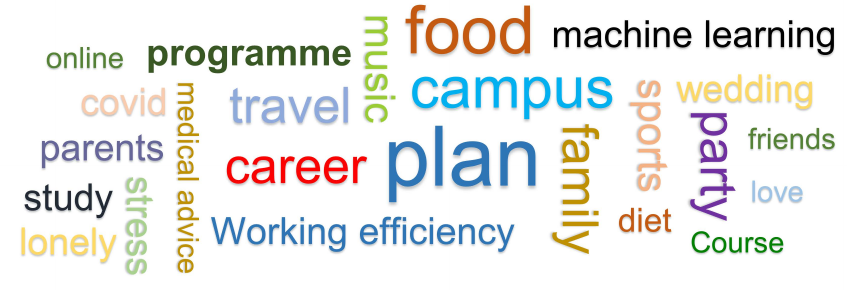}
  \caption{Keywords of the sample topics.}
  \label{words}
\end{figure}

We have randomly selected 75 specific sample topics, which we have divided into three categories: Emotional Support, Advice Acquisition, and Task-Oriented. The keywords for these topics are displayed in Figure \ref{words}. Emotional Support topics are geared towards providing comfort and guidance during challenging times, such as coping with grief, managing stress, or maintaining a positive mindset during unemployment. Advice Acquisition topics revolve around seeking suggestions and recommendations for various situations, such as planning a trip, improving time management skills, or adopting a healthier diet. Task-oriented topics involve collaborating with the chatbot to accomplish specific tasks or projects, such as planning a weekly fitness routine, organizing a charity event, or creating a budget for a home renovation. Participants have the option to choose from the provided sample topics or select their own within these three categories. We encourage participants to pick a topic that not only interests them but also aligns with their goals or passions. This approach will contribute to a more engaging and productive discussion with the chatbot.

\subsection{Procedure}
After obtaining informed consent from the participants, the experimenter began by explaining the study's purpose: to gain insight into user perceptions and interactions with the system and to identify any potential obstacles. The experimenter then guided the participants through the following steps:

Step 1: Participants were asked to select an initial topic from the provided sample topics or choose a topic of personal interest.

Step 2: Participants were instructed to engage in a conversation with the chatbot within their chosen topic.

Step 3: After the conversation, participants were requested to review it and verbally rate its quality on a scale of 0 to 100.

Step 4: Participants repeated this process twice more, selecting different topic categories each time.

During these conversations, participants were provided with topics and initiated discussions using the system. They were encouraged to freely select topics that interested them, with the intention of prompting the chatbot to provide advice and emotional support. The rating process was designed to help participants critically assess the chatbot's effectiveness in addressing their concerns and meeting their expectations. To familiarize themselves with the chatbot's functionality, participants were initially encouraged to engage in casual conversations with it twice. At the conclusion of each interview, participants were asked to reflect on their understanding of LLM-based chatbots and rate their conversations accordingly.

% For the conversations, participants received topics and started conversations via the system. They were encouraged to freely choose topics that interested them, with the aim of prompting the chatbot to provide advice and emotional support. The rating process was designed to help participants critically assess the chatbot's effectiveness in addressing their concerns and meeting their expectations.
% Initially, participants were encouraged to casually converse with the chatbot twice to familiarize themselves with its functionality. At the end of each interview, participants were asked to reflect on their understanding of the LLM-based chatbots and rate their conversations accordingly.

\subsection{Participants and Data Collection}
We recruited 14 participants from a university, consisting of 6 males and 8 females, aged between 18 and 39 years old. These participants represented a variety of academic majors, including engineering, business, psychology, computer science, pharmacy, and social science. The experiment had a duration of 40 to 60 minutes and was conducted online through a chatbot website. This website accessed the OpenAI API, utilizing the GPT-3.5-Turbo model for the conversations. Participants provided consent for their conversations to be recorded as part of the study.

\subsection{Results}
\subsubsection{Participants’ Understandings of the Large Language Model-based chatbot}

\
\\
We identified four themes from the participants’ understandings of ChatGPT from the interviews.
%人的描述可能会让机器没法理解
% 增加任务完成度，更加周全
% 机器对人的关注度

%Limited context understanding: Although ChatGPT performs well in handling many tasks, it may struggle with understanding complex or ambiguous contexts, resulting in less accurate or relevant responses.

%Biases in generation: As the model is trained on a vast amount of internet text, it might learn biases or stereotypes from the training data, which could be reflected in the generated responses.

\textbf{Lack of multi-step reasoning capabilities}: ChatGPT may not excel in tasks that demand multi-step reasoning or in-depth analysis, as its primary focus is on generating reasonably coherent short-term responses. Some participants noted that the chatbot tends to provide plausible responses without delving into a comprehensive understanding of the topic. Consequently, it often offers brief responses rather than providing a detailed analysis.

%Verbosity or repetition: ChatGPT may sometimes generate overly verbose or repetitive responses, which can affect the quality and effectiveness of the answers.

%Safety and content filtering issues: Despite some degree of content filtering, ChatGPT might still generate inappropriate or offensive responses.

%Difficulty in guiding: The quality of ChatGPT's responses can vary greatly depending on the user's prompts. Users might need to try multiple times to find the right prompt to obtain the desired answer.

%\textbf{Inconsistency}: ChatGPT may sometimes provide inconsistent responses or change its stance between different responses, which could lead to confusion for users.

\textbf{Lack of personalization}: ChatGPT may not be able to tailor its responses to individual users' preferences, backgrounds, or specific needs, which could impact user satisfaction. Some participants assume that some responses provide a brief explanation, but it's not tailored to the specific needs or preferences of the user, who might be looking for more detailed guidance or step-by-step instructions as a beginner.

%\textbf{Misinterpreting user intent}: The model might occasionally misinterpret the user's intention behind a prompt, resulting in irrelevant or unhelpful responses.

\textbf{Sensitivity to prompt phrasing}: The quality of ChatGPT's responses can be highly dependent on how the prompt is phrased, which might require users to experiment with different phrasings to get the desired output. ChatGPT provides two responses based on the slight variation in the user's prompt phrasing. For example: The first prompt, "How can I improve my communication skills?" results in a more concise response with general suggestions. However, when the user phrases the prompt as "Tips for better communication skills," ChatGPT generates a list of more specific tips. The change in phrasing has led to different forms of advice, showcasing the model's sensitivity to the way prompts are phrased.

\textbf{Absence of emotions and empathy}: ChatGPT, being an AI model, lacks genuine emotions and empathy, which could limit its ability to provide truly empathetic and emotionally appropriate responses in certain situations. For example, when a participant asks how to cope with the loss of his pet passed away. While this response offers some general suggestions for coping with the loss of a pet, it might not fully convey the depth of empathy and emotional understanding that a human would be able to provide in such a situation. 

\subsubsection{Participants’ rate of the conversation}
\
\\
Participants rate the conversation based on the chatbot's effectiveness in addressing their concerns, accomplishing tasks, and providing true empathy. There are some factors that most influence the score.

\textbf{The conversation topic is familiar to the participants or not.} We discovered that when participants are familiar with the conversation topic, they can supply more detailed prompts, leading to the chatbot generating more effective responses and receiving higher satisfaction scores. Conversely, when participants have limited knowledge of the topic, the chatbot tends to provide brief explanations, and the participants find it challenging to ask for in-depth guidance or step-by-step instructions due to their lack of background knowledge. This affects the participants' experience and results in lower satisfaction scores.

%\textbf{rational response without emotion} We found that ChatGPT is not capable of providing true empathy and emotion. It can mimic empathetic and emotional language to some extent, but does not possess genuine feelings or emotions like a human does. It lacks the personal experiences and emotions that enable humans to genuinely empathize and connect with others and results in lower satisfaction scores.
\textbf{Objective response devoid of emotion:} We found that ChatGPT cannot exhibit genuine empathy or emotions. While it can simulate empathetic and emotional language to a certain degree, it lacks the authentic feelings and personal experiences that enable humans to empathize and connect with others truly. This limitation leads to lower satisfaction scores.

\section{Technical method}
Through our interviews, we gained insights into how laypersons interact with chatbots and discovered that large language models heavily rely on prompts to generate outputs. However, humans often have limited knowledge, making it challenging to provide in-depth guidance or step-by-step instructions to chatbots due to their lack of background knowledge. This gap presents a significant obstacle in human-chatbot interactions.

\begin{figure}
  \includegraphics[width=1.0\linewidth]{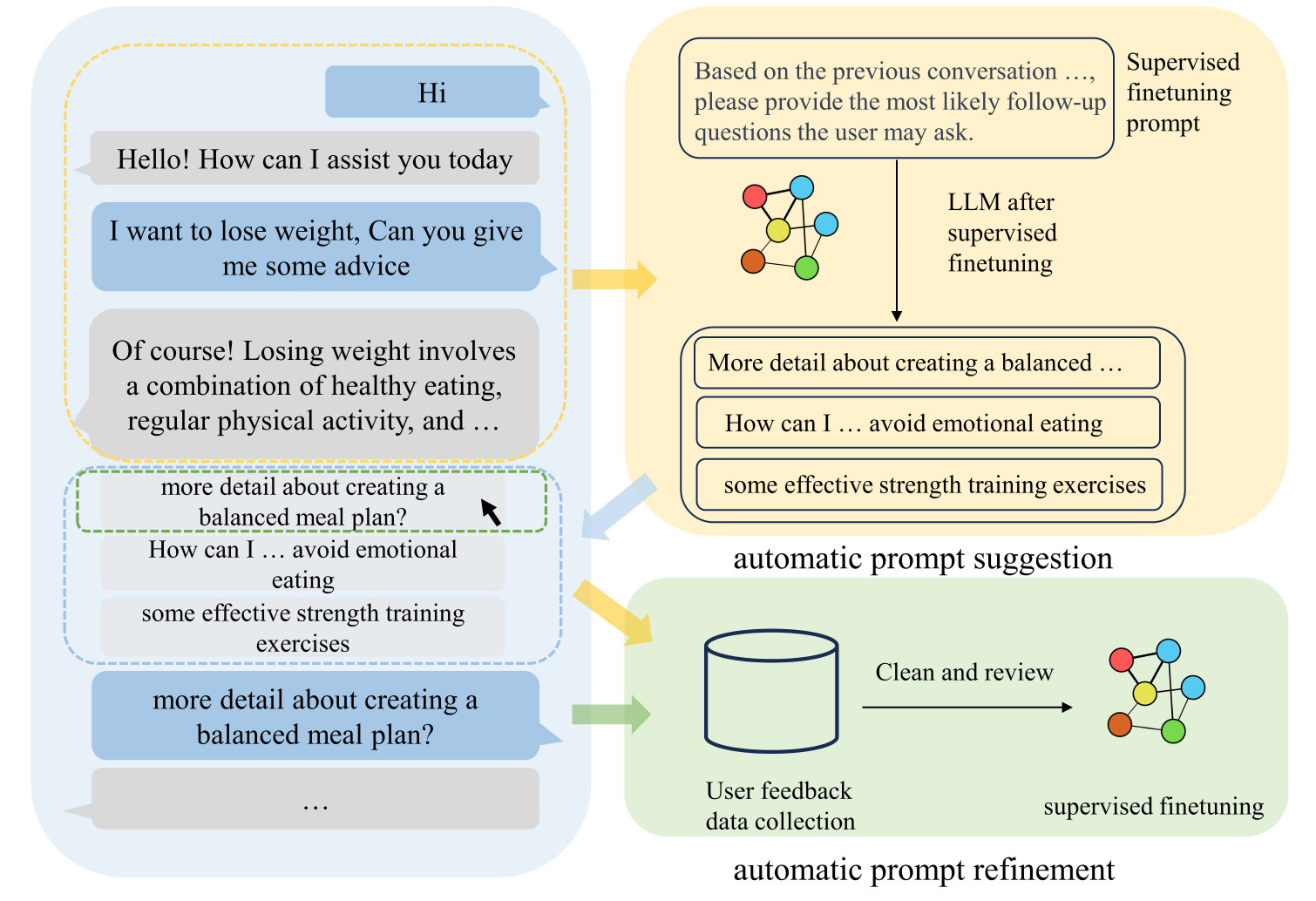}
  \caption{Workflow of PromptMind Automatic prompt suggestions generate the most suitable next prompts for users to continue dialogue with chatbots, it enables more comprehensive and contextually relevant responses. Automatic prompt refinement utilizes user feedback to improve prompt quality. Users can select from suggested prompts that are categorized as good cases for future improvement of the suggestion system.}
  \label{model}
\end{figure}

%In this section, we introduce "Prompt Your Mind," a novel prompt engineering method designed to automatically generate the next plausible prompt in a conversation for the user. This approach aims to bridge the knowledge gap and enhance the overall interaction experience between users and chatbots.

In this section, we present PromptMind a novel prompt engineering method that offers an automated way to generate the most suitable next prompt in a conversation for the user. The primary objective of this approach is to narrow the knowledge gap and elevate the overall interaction experience between users and chatbots. By leveraging PromptMind, users can enjoy a more seamless and engaging conversation with chatbots, as it intelligently generates prompts that align with their ongoing dialogue. This innovative technique acts as a bridge, enabling a smoother flow of information and enhancing the overall usability of chatbot interactions. Moreover, As illustrated in Figure \ref{model}, PromptMind improves the workflow of human-chatbot interaction by offering two main features, including 1) automatic prompt suggestion, and 2) automatic prompt refinement.

\subsection{Automatic Prompt Suggestion}
We aim to address the challenge of constructing effective prompts for human-chatbot interaction through our prompt suggestion engine. Our system utilizes a smaller foundation model called LLaMA (7B parameters)\cite{touvron2023llama} and applies supervised fine-tuning to develop the Automatic Prompt Suggestion model.

%, we can create a dataset that is representative of the target users and provides a comprehensive understanding of their needs and preferences. 
To enhance our training dataset, we employ few-shot prompting techniques to expand the data generated by ChatGPT. Initially, we utilize multi-turn conversation data from the interview study with high satisfaction scores (>85) as a starting point. However, we take precautions to protect sensitive information by removing or replacing personal identifiers like names, addresses, and phone numbers. Subsequently, we expand this dataset using ChatGPT. Before creating the supervised fine-tuning dataset, we define the scope of the conversation topics and carefully select and balance the conversation topics. During inference, we generate relevant text data by providing a few demonstration conversations as conditioning. We use ChatGPT3.5 with a temperature parameter set to 0.9 to encourage creative output. The generated responses are thoroughly reviewed and edited to ensure their accuracy and relevance to our research. This edited output serves as a starting point for generating additional data to expand our dataset.

Once we have completed the supervised fine-tuning process on the LLaMA model using the expanded conversation dataset, we can infer the model to obtain automatic prompt suggestions. Specifically, at each round of conversation, the system uses the previous conversation as input to infer the model (Due to the input limitations of the language model, it takes into account the four preceding rounds of conversation). and generates suggested prompts based on this input, helping the user further express their intention.

By using the preceding conversation as context, the model can better understand the user's current needs and the context of the conversation. Consequently, the generated extended prompts will more accurately reflect the user's intent and provide targeted and coherent responses. This contextual approach improves the fluency and comprehensibility of the dialogue and generates effective prompts for chatbot interactions. In summary, by leveraging the preceding conversation information to generate extended prompts for user inference, we enhance our understanding of the user's needs and deliver more accurate and targeted responses.

\subsection{Automatic Prompt Refinement}
The automatic prompt refinement process is a dynamic user-driven feedback loop that empowers the continuous enhancement and optimization of our prompt suggestion engine. When we provide users with suggested prompts, they have the option to click on a button associated with each prompt. Clicking on a suggested prompt triggers an automatic process that sets the chosen prompt as the input for the chatbot. This indicates that the chosen prompt has met the user's expectations and is considered a good case. We utilize these cases to improve and enhance the features of our suggestion system in future iterations.
By collecting user feedback and identifying the prompts that successfully assist users in achieving their desired outcomes, we can refine and optimize the prompt suggestion engine. This iterative process allows us to continuously learn from user interactions and improve the overall user experience.
Through this user-driven prompt refinement approach, we aim to ensure that the suggested prompts are highly relevant and effective in helping users express their intentions accurately. This ongoing improvement helps create a more seamless and satisfying human-chatbot interaction.
%Promptify improves the workflow of text-to-image generation by offering three main features, including 1) automatic prompt exten- sion and suggestion, 2) image layout and clustering by similarity, and 3) automatic prompt refinement suggestions. By integrating these features, Promptify establishes a feedback loop that enables users to generate high-quality images from initial prompts, examine how the model interprets their prompt, and receive suggestions to edit the initial prompts for enhancing output. Figure 2 illustrates the user workflow using Promptify. Next, we will introduce individ- ual features and associate them with the design goals outlined in Section 4 to underscore their design rationale.
\section{Evaluation}
After demonstrating our PromptMind method, we sought to answer Our third research question:
RQ3: How effective is the PromptMind approach in enhancing the quality of human-chatbot interactions?

\subsection{Experimental design}

To validate the effectiveness of PromptMind, we conducted a within-subject study to compare our system with the original prompt. The experimental stimuli consisted of three types of dialogue topics, including emotional support, advice acquisition, and task-oriented topics. The prompt editing conditions have two levels: Original Prompt and PromptMind. Emotional support includes expressions of care, concern, love, and interest, and also encompasses helping distressed others work through their upset by listening to, empathizing with, legitimizing, and actively exploring their feelings \cite{burleson2003emotional,reblin2008social}. The advice acquisition topics include information and advice seeking. Task-oriented topics aim to complete certain tasks for users in a specific domain, such as restaurant recommendation, and location query, which makes it valuable for real-world business \cite{zhang2020recent,liu2017rubystar}. Overall, each participant in this study needs to complete an experiment with 2 prompt editing conditions, and each has 3 tasks.

The experiment was conducted online via an online chatbot website that accesses the OpenAI API to utilize the GPT-3.5-Turbo model.

%%The study aimed to evaluate the impact of two different prompt editing conditions on the conversation. The independent variable in this experiment was the Prompt Editing Conditions with two levels: Original Prompt and PromptMind.
%%We used the same three categories of topics (Emotional Support, Advice Acquisition, and Task-Oriented) as in the previous interview study. Participants could choose from the provided sample topics or select their own within these three categories. We encouraged them to pick a topic that interested them and aligned with their goals or passions to ensure a more engaging and productive discussion with the chatbot.

\subsection{Experimental procedure}

In this experiment, participants needed to complete 6 tasks (3 types $\times$ 2 Prompt Editing Conditions). They are asked to choose one task from the three categories of tasks to start their experiment. The whole experiment last approximately 60-75 minutes. %The experimental procedure is as follows: (最好加一张流程图)

%%Here is the procedure we followed in the user study:

Step 1: Introduction and Consent: Participants were introduced to the objective of the experiment and asked for their consent to participate.

Step 2: Topic Selection: Participants selected an initial topic from the provided sample topics or chose a topic of personal interest.

Step 3: Prompt Editing Condition: Participants were assigned one of the Prompt Editing Conditions: Original Prompt or PromptMind.

Step 4: Conversation Engagement: Participants engaged in a conversation with the chatbot within the chosen topic, utilizing the assigned Prompt Editing Condition.

Step 5: Conversation Review and Questionnaire: After the conversation, participants reviewed the conversation transcript and completed the post-questionnaires. These questionnaires aimed to gather feedback on their experience and evaluate the effectiveness of the conversation under the assigned Prompt Editing Condition.

Step 6: Repeat Process: The above steps (2-5) were repeated twice more with different topics to capture a wider range of experiences and opinions.

%%By conducting this user study, we aim to gather valuable insights into the impact of different prompt editing conditions on the conversation quality. The collected feedback will help us refine and improve the system accordingly.

\subsection{Participant}
24 participants (14 females and 10 males), recruited from a university, took part in the experiments. They aged from 19 to 32 years old and had a normal cognitive ability. The conversations and the questionnaires were recorded with the participant's consent.

% who completed the survey in approximately 60-75 minutes and received compensation.
 %Each participant completed 6 questionnaires (3 types $\times$ 2 Prompt Editing Conditions). We conducted the study online via an online chatbot website that accesses the OpenAI API to utilize the GPT-3.5-Turbo model. The conversations and the questionnaires were recorded with the participant's consent.

\subsection{Measures}
In this study, we examined three topic categories (Emotional Support, Advice Acquisition, and Task-Oriented topics) to evaluate two distinct Prompt Editing Conditions: Original Prompt and PromptMind. We assessed the social presence and task usability for both emotional support and advice acquisition tasks. Meanwhile, for the Task-Oriented category, we focused on measuring users' perceived workload.

%We measured social presence by questionnaires. The social presence questionnaire was based on Harms and
%Biocca’s \cite{harms2004internal} questionnaire, which conceptualizes social presence in six dimensions: co-presence, the degree to which the observer believes s/he is not alone; attentional allocation, the amount of attention the user allocates to and receives from an interactant; perceived message understanding, the ability of the user to understand the message from the interactant; perceived affective understanding, the user’s ability to understand the interactant’s emotional and attitudinal states; perceived affective interdependence, the extent to which the user’s emotional and attitudinal state affects and is affected by the interactant’s emotional and attitudinal states; and perceived behavioural interdependence, the extent to which the user’s behaviour affects and is affected by the interactant’s behaviour. The social presence questionnaire was translated to the subjects’ native language. We selected two items for each dimension that would be adequate for children (see Table 1). Subjects were asked to express their agreement or disagreement regarding each item on a five-point Likert scale (zero means “totally disagree” and five “totally agree”).

\begin{itemize}
%\vspace{-0.2cm}
  \item [1)] Social presence\\
The social presence is measured by the social presence questionnaire designed by Harms and Biocca's \cite{harms2004internal}. It conceptualizes social presence across 4 dimensions: co-presence, perceived message understanding, perceived affective understanding, and perceived behavioral interdependence. The questionnaire was translated into the participants' native language. Participants were asked to indicate their level of agreement or disagreement for each item using a five-point Likert scale, where zero represents "totally disagree" and five signifies "totally agree".
   \item [2)] Workload\\
The users' perceived workload is tested by the task load index (NASA-TLX) \cite{hart2006nasa}, which was developed by NASA. It consists of six subscales: mental demand, physical demand, temporal demand, performance, effort, and frustration. Each subscale is rated on a 21-point scale, and the ratings are combined to calculate a total workload score. 
   \item [3)] System usability\\
The system usability is tested by the post-study system usability questionnaire (PSSUQ) \cite{lewis1992psychometric}. It consists of 9 items and each item rates on a 7-point Likert scale, ranging from 1 (strongly disagree) to 7 (strongly agree). The questionnaire aims to provide a comprehensive understanding of the user's experience, such as ease of use, efficiency, effectiveness, and satisfaction with the system.
\end{itemize}

\section{Result}

%Here's the corrected and clarified version of your text:
\subsection{Result of social presence}
\begin{figure}
  \includegraphics[width=\textwidth]{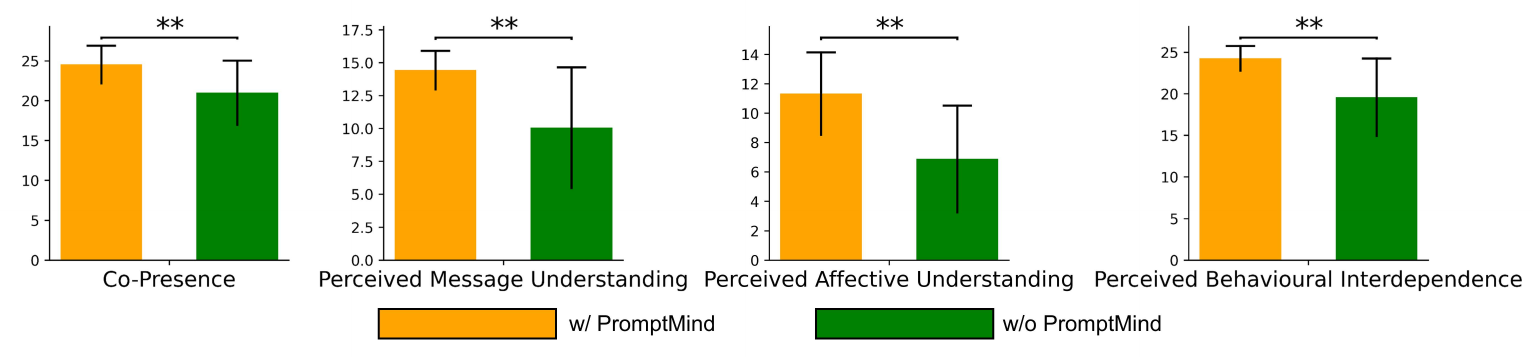}
  \caption{Results of the total scores of the ratings from the comparative evaluation in terms of social presence. PromptMind receives significant improvements compared with the baseline. Error bars represent standard deviation. * indicates $p <0.05$, and ** indicates $p <0.01$.}
  \label{res_pssuq}
\end{figure}

As depicted in Figure \ref{res_pssuq}, PromptMind consistently received higher average scores across all ratings in terms of social presence ($Z=-1.207$, $p=0.004$). Specifically, within the dimension of co-presence, the refined and extended prompts generated by PromptMind facilitated improved interaction between users and the agent, enhancing their awareness of each other's presence. It is likely that these refined prompts played a pivotal role in capturing users' attention, which contributed to these positive outcomes. In the category of perceived message understanding, a larger number of users reported a clearer understanding of the agent's thought process when using PromptMind. Additionally, most users affirmed that their own thoughts were effectively conveyed to PromptMind. This trend also extended to the dimension of perceived affective understanding. For instance, users sometimes expressed surprise at how well the refined prompts aligned with their thoughts. In the final dimension, perceived behavioral interdependence, it became evident that the improved understanding scores led to quicker responses during conversations. This suggests that our PromptMind system has the potential to enhance dialogue efficiency. Furthermore, a detailed analysis using a Mann-Whitney Test revealed that our system significantly bridged the understanding gap between the chatbot and the user. The differences between PromptMind and the original prompt were found to be significant in terms of co-presence ($Z=-1.505$, $p=0.0005$), perceived message understanding ($Z=-3.001, p=0.0002$), and perceived affective understanding ($Z=-1.584, p=0.0004$). These results indicate that the PromptMind system provides social benefits for users engaging with the agent and fosters deeper connections with the system.

\subsection{Result of workload}

\begin{figure}
  \includegraphics[width=\textwidth]{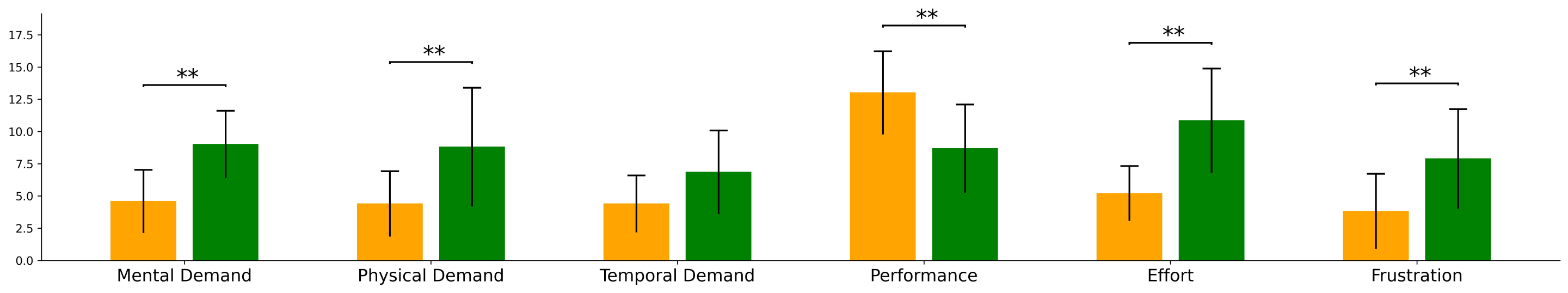}
  \caption{Results of the total scores of the ratings from the comparative evaluation in terms of TLX. PromptMind receives significant improvements compared with the baseline. Error bars represent standard deviation. * indicates $p <0.05$, and ** indicates $p <0.01$.}
  \label{res_tlx}
\end{figure}

Figure \ref{res_tlx} illustrates the task load demands within the context of information-seeking conversations. Across all dimensions, when PromptMind is employed, there is a consistent reduction in task load demands compared to when PromptMind is not used. Notably, statistically significant differences are observed in five dimensions: mental demand ($Z=1.829, p<0.0001$), physical demand ($Z=1.765, p=0.0004$), performance ($Z=-1.356, p<0.0001$), effort ($Z=2.688, p<0.0001$), and frustration ($Z=1.414, p=0.0004$), with the exception of temporal demand. It has been observed that users tend to prefer engaging in longer conversations with the chatbot when PromptMind is utilized. This preference can be attributed to several factors, including improved affective understanding, which enhances the social connection between the chatbot and users. Additionally, PromptMind generates more relevant and subconscious questions that align with users' interests, potentially leading to longer conversations. In this context, the duration of the conversations naturally tends to increase. In summary, the utilization of PromptMind leads to an overall reduction in task load demands. This underscores the effectiveness of PromptMind in enhancing the efficiency of information-seeking interactions.

\begin{figure}
  \includegraphics[width=0.6\linewidth]{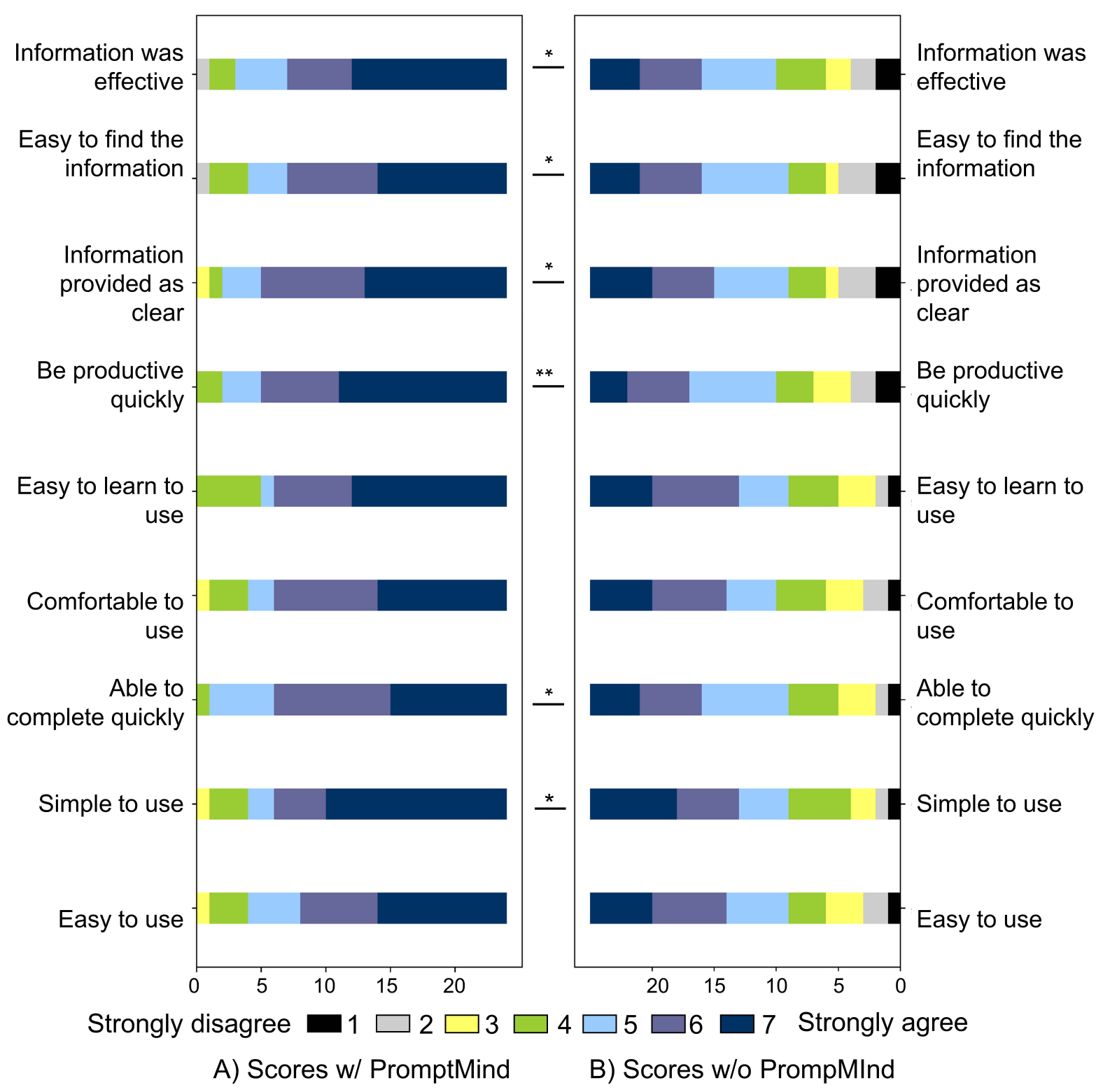}
  \caption{Results of the total scores of the ratings from the comparative evaluation in terms of PSSUQ. PromptMind receives significant improvements compared with the baseline. Error bars represent standard deviation. * indicates $p <0.05$, and ** indicates $p <0.01$.}
  \label{res_pss}
\end{figure}

%  5.95|1.2031209415515967-5.1875|1.549
% 6.25|1.2196310917650468-5.5625|1.3678
%  6.05|1.16081867662439-5.125|1.5761

\subsection{Result of system usability}

We conducted an analysis of user feedback, utilizing questionnaire responses and interview data to assess the user experience. Subjective ratings of the user experience are presented in Figure \ref{res_pss}, with the statistical results from a Mann-Whitney Test indicated by asterisks (* for $p<0.05$ and ** for $p<0.01$). Initially, participants found PromptMind's refined prompts to be useful, scoring an average of 5.95 (STD=1.20), and perceived them as easier to use than the prompts without PromptMind, which received an average score of 5.19 (STD=1.55). There was a marginally significant difference in terms of ease of use ($Z=-1.221,p=0.04$). Additionally, PromptMind enabled users to locate the required information more swiftly. This was particularly evident in the dimension of productivity, where the score was 6.4 (STD=0.86) with PromptMind compared to 4.63 (STD=1.65) in the baseline. A significant difference was observed ($Z=-2.063,p=0.0005$). In the dimension of completing quickly, PromptMind scored 6.10 (STD=1.30), while the baseline achieved a score of 4.81 (STD=1.70). A marginally significant difference was found ($Z=-0.861, p=0.035$). In summary, PromptMind not only enhances information-seeking efficiency but also improves the overall user experience.

% 6.4|0.8602325267042626-4.625|1.6535
%  6.1|1.3-4.8125|1.703626
\section{summary}

Our study has shed light on the current practices and limitations of how users interact with ChatGPT. It has become evident that while ChatGPT is an impressive language model, it still faces certain challenges. These limitations include a lack of multi-step reasoning capabilities, limited personalization, sensitivity to prompt phrasing, and the absence of emotions and empathy.

%To address these limitations and improve the quality of human-chatbot interactions, a novel approach called PromptMind was proposed. This approach focuses on automatically giving the most suitable next prompts to generate improved chatbot responses. The approach offers prompt suggestions to help users continue conversations even when they lack specific knowledge or guidance. It also provides prompt refinement suggestions based on user feedback, allowing for enhanced prompt quality and a more tailored interaction experience.

To address the limitations and enhance the quality of human-chatbot interactions, a novel approach called PromptMind has been proposed. This method concentrates on automatically providing the most suitable next prompts to generate improved chatbot responses. The approach offers prompt suggestions that assist users in continuing conversations, even when they lack specific knowledge or direction. Additionally, PromptMind incorporates prompt refinement suggestions based on user feedback, allowing for improved prompt quality and a more personalized interaction experience.

To evaluate the effectiveness of the PromptMind approach in enhancing human-chatbot interactions, a study was conducted. Three topic categories were examined: Emotional Support, Advice Acquisition, and Task-Oriented discussions. Two prompt editing conditions were compared: Original Prompt and PromptMind. The results of the evaluation showed significant improvements across different metrics. In terms of social presence and task usability, the PromptMind approach outperformed the Original Prompt condition. Additionally, users reported reduced perceived workload when using the PromptMind approach.

Overall, the PromptMind approach offers a promising solution for improving human-chatbot interactions. By addressing the limitations of current practices, it provides users with a more personalized and engaging experience. With continued advancements and refinements, this approach has the potential to revolutionize the way we interact with chatbots, opening up new possibilities for meaningful and effective communication.

% \section{Discussion}
% \subsection{Design Implications and Limitations}
% \textbf{we improve the social connection but can }
% \textbf{}
% \subsection{future work}

% \section{Acknowledgments}

% \section{Appendices}

\bibliographystyle{ACM-Reference-Format}
\bibliography{sample-base}

\appendix









\end{document}